\documentclass[sigconf, nonacm]{acmart}
\usepackage{graphicx}

\usepackage{amsmath, amsthm, amssymb}
\usepackage{algorithm}
\usepackage{algpseudocode}
\usepackage{booktabs, subcaption, threeparttable}
\usepackage{caption, placeins}
\usepackage{stfloats}
\usepackage{comment}
\usepackage{xcolor, soul}
\usepackage{hyphenat}
\usepackage{pifont}
\usepackage{url}

\theoremstyle{definition}

\newcommand{\crvs}[1]{\textcolor{black}{#1}}

\begin{document}
\title{Understanding Domain-Aware Distribution Alignment in Budgeted Entity Matching}

\author{Nicholas Pulsone}
\affiliation{%
  \institution{Worcester Polytechnic Institute}
  \city{Worcester}
  \state{MA}
}
\email{nbpulsone@wpi.edu}

\author{Gregory Goren}
\affiliation{%
  \state{Israel}
}
\email{grggoren@gmail.com}

\author{Roee Shraga}
\affiliation{%
  \institution{Worcester Polytechnic Institute}
  \city{Worcester}
  \state{MA}
}
\email{rshraga@wpi.edu}

\begin{abstract}
Entity Matching (EM) is a core operation in the data integration pipeline, where records from different sources are compared to determine whether they refer to the same real-world entity. Recent work has incorporated domain information and low-resource learning techniques to better adapt EM systems to realistic settings. While these approaches have demonstrated strong performance, it remains unclear how they behave under varying data constraints and levels of supervision in practice. In this paper,\footnote{This is the extended version of a workshop
paper to appear in
TaDa 2026~\cite{pulsone2026dist}.} we investigate a state-of-the-art method for low-resource, domain-aware EM--\emph{BEACON}~\cite{pulsone2026beacontr}--and study how its performance is affected by different algorithmic choices and data availability conditions. We conduct a series of targeted experiments to evaluate these variations, providing deeper insight into the role of distribution alignment and the behavior of the BEACON framework.
\end{abstract}

\maketitle

\section{Introduction} \label{sec:introduction}

Entity Matching (EM) is a fundamental operation in tabular data management. Commonly used for data integration, deduplication, and data cleaning, EM aims to determine whether two records refer to the same real-world entity. Historically, EM has proven to be a challenging research problem~\cite{fellegi1969theory, winkler1990string, hernandez1995merge, Bilenko03}, but with increasingly effective approaches emerging in recent years~\cite{ebraheem2017deeper, mudgal2018deep, li2020ditto, peeters2023llm}. Today, state-of-the-art EM systems typically fine-tune large Pre-trained Language Models (PLMs)~\cite{li2020ditto, hiergat} or Large Language Models (LLMs)~\cite{peeters2023llm, steiner2025ftLLM} on task-specific datasets. While highly effective, these approaches face practical limitations, including high training and inference costs. A key bottleneck is the scarcity of labeled data for new matching tasks. Consequently, recent work has explored low-resource and budget-aware EM methods that aim to achieve strong performance with limited labeled data~\cite{kasai2019low, genossar2023battleship}.

In practice, however, practitioners often have access to labeled data from multiple \emph{domains}, which can provide valuable signals for learning under limited budgets. To address this setting, BEACON~\cite{pulsone2026beacontr} introduces the problem of \emph{Budget-Aware Entity Matching Across Domains} (EMAD), where the goal is to train an effective EM model for a target domain by selectively leveraging labeled data from other domains under a fixed annotation budget. BEACON employs distribution-aware sampling strategies to identify informative out-of-domain samples. In particular, the Train–Validation Distribution Fitting (TVDF) method selects samples that improve alignment between the training distribution and a target (validation) distribution. Although BEACON demonstrates strong empirical performance, it remains unclear why distribution-aware methods such as TVDF are effective. In this work, we take a closer look at TVDF and investigate the role of distribution alignment in budgeted EM. We study how TVDF behaves under varying practical conditions and explore extensions that can improve both performance and interpretability.

To this end, our contributions are as follows\footnote{Code and data 
available: \url{https://github.com/nbpulsone/BEACON/tree/dist-alignment}}:
\begin{itemize}
\setlength\itemsep{0.5em}

\item \textit{Label Availability.} We analyze the behavior of TVDF under varying levels of label availability, including settings with limited or missing in-domain and out-of-domain labels.

\item \textit{Alternative Domain Representations and Selection Criteria.} We investigate alternative representations of EM data distributions and evaluate their impact on distribution alignment and downstream performance.

\item \textit{Domain-Agnostic Distribution Alignment.} We isolate the distribution-alignment component of TVDF through controlled downsampling, showing that distribution-aware selection can reduce training data requirements while mitigating performance loss without explicit domain structure.

\end{itemize}

The rest of the paper is organized as follows. Section~\ref{sec:preliminaries} reviews the necessary background on EMAD and budgeted sampling. Section~\ref{sec:distribution} presents key insights into distribution alignment with a focus on TVDF. Section~\ref{sec:experiments} reports experimental results and analysis across multiple settings, and Section~\ref{sec:conclusion} concludes the paper.

\section{Preliminaries} \label{sec:preliminaries}

\noindent\textbf{Entity Matching (EM).}
EM is the task of determining whether two records refer to the same real-world entity. EM typically consists of two stages: (i) \emph{blocking}, which reduces the quadratic space of candidate pairs using computationally efficient techniques, and (ii) \emph{matching}, which applies a more expressive model to classify candidate pairs. In this work, we focus on the matching phase and assume as input the output of an effective blocking procedure.

Each candidate pair (or sample) consists of two serialized records $r_1$ and $r_2$. Following standard PLM-based EM approaches~\cite{li2020ditto, genossar2023battleship}, the input is represented as:
\[
\texttt{[CLS]} \; r_1 \; \texttt{[SEP]} \; r_2
\]
where the \texttt{[CLS]} token encodes a summary representation of the pair and \texttt{[SEP]} separates the records. A model is trained to predict a binary label indicating whether the records match.

\vspace{0.5em}
\noindent\textbf{Budgeted Entity Matching Across Domains (EMAD).}
BEACON~\cite{pulsone2026beacontr} introduces the EMAD setting, where candidate pairs $N$ are partitioned into domains $\mathcal{D}$:
\[
N = n_1 \cup n_2 \cup \dots \cup n_j, \quad n_i \cap n_k = \emptyset \;\; \text{for } i \ne k.
\]
For each domain $i \in \mathcal{D}$, the goal is to construct a domain-specific training set $X_i$ under a fixed annotation budget $\beta$. The training set may include both in-domain \crvs{($n_i$)} and out-of-domain \crvs{($S_i^{\text{out}}$)} samples:
\[
X_i = n_i^{*} \cup S_i^{\text{out}}, \quad |X_i| = \beta, \quad S_i^{\text{out}} \subseteq (N \setminus n_i),
\]
where $n_i^{*}$ denotes a (possibly sub or over-sampled) subset of in-domain data. The objective is to select $X_i$ such that the resulting model achieves strong matching performance on domain $i$.

\vspace{0.5em}
\noindent\textbf{BEACON.}
BEACON~\cite{pulsone2026beacontr} is a framework for EMAD that leverages distribution-aware sampling to select informative out-of-domain samples. It operates over embedding representations of candidate pairs, obtained from a backbone language model via the \texttt{[CLS]} token. These embeddings are used to guide sample selection, favoring candidates that align with the target domain distribution.

A key component of BEACON is its dynamic training procedure, in which embeddings are periodically recomputed as the model is fine-tuned. This allows the sampling process to adapt to the evolving representation space, resulting in progressively better alignment between the training data and the target domain.

\section{Distribution Alignment} \label{sec:distribution}

A natural way to leverage embeddings for out-of-domain selection in EMAD is a nearest-neighbor (NN) approach: given a target-domain centroid $\mu_i$, select out-of-domain samples closest to $\mu_i$ under a similarity metric (e.g., cosine similarity). While intuitive, this strategy preserves the original training distribution and does not correct for cases where the in-domain data is sparse or unrepresentative. In practice, domain sizes can vary significantly (e.g., WDC exhibits a highly skewed domain distribution), limiting the effectiveness of such local selection strategies.

To address this, BEACON~\cite{pulsone2026beacontr} introduces the \emph{Train–Validation Distribution Fitting} (TVDF) method, which leverages a proxy (validation) distribution to guide sample selection. 
Rather than preserving the training distribution, TVDF selects out-of-domain samples that improve alignment between the training centroid $\mu_i$ and a proxy centroid $\mu_{\text{val}_i}$. Formally, given a set of candidates $A$, TVDF selects:
\[
\text{TVDF}_k(A, \mu_i, \mu_{\text{val}_i}) =
\operatorname*{Top\text{-}k}_{x \in A}
\big[
\cos(\mu_{n_i \cup x}, \mu_{\text{val}_i}) -
\cos(\mu_i, \mu_{\text{val}_i})
\big].
\]

\crvs{where $\cos$ is the cosine-similarity operator and $\mu_{n_i \cup x}$ is the centroid of the in-domain data when sample $x$ is added}.

While effective, TVDF relies on several simplifying assumptions. First, it assumes that no label information is available during selection, which may not always hold in practice. Second, the domain population is modeled by its centroid, and distribution fitting is assessed according to centroid alignment. Finally, TVDF assumes that domain structure is explicitly defined and available to guide sampling. We next examine these assumptions.


\vspace{0.5em}
\noindent\textbf{Label-Aware Sampling.}
While TVDF is inherently unsupervised, it can naturally incorporate additional signal when label information is available. In a label-aware setting, embeddings may be partitioned into positive (matching) and negative (non-matching) subsets, enabling alignment of class-specific distributions rather than treating the data as a single homogeneous pool. Thus, separate centroids can be maintained for positive and negative samples, and out-of-domain candidates can be selected to improve alignment with each class-specific validation distribution.

This formulation allows the sampling process to better respect class imbalance and structural differences between matches and non-matches, which are often pronounced in EM tasks. When out-of-domain labels are also available, selection can be performed independently over labeled pools, ensuring that both positive and negative samples are appropriately represented. 

It should be noted that when including out-of-domain label information in the sampling process, we depart from the original motivation of acquiring annotations via a budget--since these samples are already labeled. However, out-of-domain sampling can remain practically relevant for reducing training cost, improving sample efficiency, or leveraging partially labeled external data sources.

\vspace{0.5em}
\noindent\textbf{Alternative Domain Representations and Selection Criteria.}
TVDF represents distributions using centroids, which capture central tendency but ignore higher-order structure. Alternative representations can provide additional information about the embedding space. One option (TVMed) is to use \crvs{the medoid for a given domain $i$ ($M_i$)}, defined as the most central sample in that domain:
\[
M_i = \arg\min_{x \in n_i} \sum_{y \in n_i} d(x, y),
\]
where $d$ represents the cosine distance function. This representation is arguably more robust to outliers, however it is computationally more expensive. Another extension incorporates variance, capturing the spread of the distribution. By jointly considering centroid and variance alignment, TVDF can model not only the central tendency of a domain distribution, but also aspects of its spread and density in the embedding space. This additional information may allow for more accurate sample selection by encouraging the training distribution to better match the structure of the proxy distribution:
\[
\text{TVDF}_{\text{VAR}} =
\operatorname*{Top\text{-}k}_{x \in A}
\left[
\Delta \cos(\mu_i, x)
+
\gamma \, \Delta \cos(\sigma^2_i, x)
\right],
\]
\[
\Delta \cos(\mu_i, x)
=
\cos(\mu_{n_i \cup x}, \mu_{\text{val}_i})
-
\cos(\mu_i, \mu_{\text{val}_i}),
\]
\[
\Delta \cos(\sigma^2_i, x)
=
\cos(\sigma^2_{n_i \cup x}, \sigma^2_{\text{val}_i})
-
\cos(\sigma^2_i, \sigma^2_{\text{val}_i}),
\]
\crvs{where $\sigma_i^2$ and $\sigma_{\text{val}_i}^2$ denote the variance representations of the training and validation data of domain $i$, respectively, $\sigma_{n_i \cup x}^2$ denotes the variance of the training data with sample $x$ added,} and $\gamma$ controls the contribution of variance alignment. Larger values of $\gamma$ place greater emphasis on matching the variance structure than the centroids of the proxy distribution during sample selection.

As a final alternative, we consider a coverage-based method (TVCoverage) that represents a domain using the full embedding space rather than a centroid summary. Instead of measuring alignment through centroid similarity, this approach measures how well the training embedding space geometrically covers the proxy/validation distribution. Specifically, the coverage \crvs{$\mathcal{C}$} is defined as:
\[
\mathcal{C}(n_i, v_i)
= \frac{1}{|v_i|} \sum_{z \in v_i}
\min_{y \in n_i} d(z, y),
\]
where each validation embedding $z \in v_i$ is assigned the distance to its nearest training embedding. Thus, minimizing $\mathcal{C}(n_i, v_i)$ means increasing coverage, so selection will favor candidate samples that reduce this quantity for a given domain $i$. Unlike centroid-based methods, TVCoverage accounts for the geometric structure of the embedding space and can capture multi-modal distributions beyond central tendency. However, this increased representational complexity comes at the cost of lower computational efficiency.

These variants aim to better approximate the true data distribution, however also 
introduce additional complexity.

\vspace{0.5em}
\noindent\textbf{Domain-Agnostic Downsampling.}
Finally, we consider whether distribution alignment remains effective without explicit domain structure. To this end, we adapt our framework to a domain-agnostic setting, where the goal is to \emph{downsample} the training data to better match a target (proxy) distribution, rather than selecting samples across predefined domains. This setup is motivated by two practical considerations. First, it allows us to study whether aligning the training distribution to a proxy distribution remains beneficial outside the standard BEACON setting, where data is explicitly partitioned into domains. In this setting, distribution alignment is instead used to identify and remove samples that appear noisy or poorly aligned with the target distribution. Second, it provides a principled approach for reducing the size of the training data in scenarios where training cost or labeling availability impose practical limitations. Given a fixed downsampling ratio, we compare:


\noindent $\bullet$ \textbf{BASE:} Training on the full dataset.\\
\noindent $\bullet$ \textbf{RANDOM:} Uniform random downsampling.\\
\noindent $\bullet$ \textbf{Nearest to Centroid (NC):} Selecting samples closest to the centroid of the proxy/validation set.\\
\noindent $\bullet$ \textbf{Nearest Neighbor (NN):} Selecting samples that are closest to any point in the proxy/validation set.\\
\noindent $\bullet$ \textbf{TVDF:} Selecting samples that improve alignment between the training and proxy/validation distributions.

These methods provide a principled framework for evaluating how TVDF-based distribution-aware selection behaves in a non-EMAD setting. 
Broadly, they provide insight into the role of distribution structure in EM independent of domain-specific assumptions.

\section{Experimental Insights} \label{sec:experiments}

We present a series of 
experiments to better understand the role of distribution alignment in TVDF (Section~\ref{sec:distribution}). Experiments are conducted on the WDC Multi-Dimensional Entity Matching Benchmark~\cite{peeters2023wdc}, focusing on the 50\% corner-case, 50\% seen-entities setting. We partition the data by product category, yielding 11 domains for the EMAD experiments. Unless otherwise stated, settings follow BEACON~\cite{pulsone2026beacontr}, including budgets ranging from 1k to 10k\crvs{, and a  RoBERTa~\cite{liu2019roberta} backbone PLM.} 

Our comprehensive results are shown in Tables~\ref{tab:label_aware} and ~\ref{tab:domain_agnostic}. Tables~\ref{tab:label_aware_macro} and ~\ref{tab:representations_macro} depict macro F1 results for the label-aware and domain representation variants -- i.e., F1 scores averaged uniformly across domains for selected budget settings. Tables~\ref{tab:label_aware_weighted} and ~\ref{tab:representations_weighted} present the corresponding weighted F1 results for the label-aware and domain representation variants, where F1 scores are averaged across domains and weighted by the number of samples in each domain.

\subsection{Label-Aware Experiments} \label{sub:label_aware}

\begin{table*}
  \centering
  \caption{Label-Aware Results (left) and Alternative Domain Representations and Selection Criteria Results (right).}
  \label{tab:label_aware}
  \label{tab:representations}
  \footnotesize
  \makebox[\linewidth][c]{
  \begin{subtable}[t]{0.25\linewidth}
    \centering
    \caption{Macro F1 Results}
    \label{tab:label_aware_macro}
    \resizebox{\linewidth}{!}{\begin{tabular}{lcccc}
\toprule
 Method & 5.0k & 10.0k & Mean & SD \\
\midrule
TVDF & \textbf{0.739} & \textbf{0.759} & \textbf{0.716} & 0.049 \\
TVDF (ID) & 0.714 & 0.716 & 0.696 & 0.040 \\
TVDF (OOD) & \underline{0.719} & \underline{0.732} & \underline{0.700} & 0.045 \\
TVDF (ID + OOD) & 0.713 & 0.724 & 0.691 & 0.056 \\
\bottomrule
\end{tabular}
}
  \end{subtable}\hfill
  \begin{subtable}[t]{0.25\linewidth}
      \centering
      \caption{Weighted F1 Results}
      \label{tab:label_aware_weighted}
      \resizebox{\linewidth}{!}{\begin{tabular}{lcccc}
\toprule
Method & 5.0k & 10.0k & Mean & SD \\
\midrule
TVDF & \textbf{0.742} & \textbf{0.749} & \textbf{0.719} & 0.051 \\
TVDF (ID) & \textbf{0.742} & 0.745 & 0.714 & 0.058 \\
TVDF (OOD) & \textbf{0.742} & \underline{0.746} & \underline{0.715} & 0.056 \\
TVDF (ID + OOD) & \underline{0.736} & 0.738 & 0.710 & 0.055 \\
\bottomrule
\end{tabular}}
  \end{subtable}\hfill
    \begin{subtable}[t]{0.24\linewidth}
    \centering
    \caption{Macro F1 Results}
    \label{tab:representations_macro}
    \resizebox{.95\linewidth}{!}{\scalebox{0.8}{\begin{tabular}{lcccc}
\toprule
Method & 5.0k & 10.0k & Mean & SD \\
\midrule
TVDF & \textbf{0.739} & \textbf{0.759} & \textbf{0.716} & 0.049 \\
TVMed & 0.715 & \underline{0.742} & 0.696 & 0.056 \\
TVDF$_{\text{VAR}}$ & 0.698 & 0.719 & 0.683 & 0.041 \\
TVCoverage & \underline{0.721} & 0.739 & \underline{0.711} & 0.049\\
\bottomrule
\end{tabular}}}
  \end{subtable}\hfill
  \begin{subtable}[t]{0.24\linewidth}
      \centering
      \caption{Weighted F1 Results}
      \label{tab:representations_weighted}
      \resizebox{.95\linewidth}{!}{\begin{tabular}{lcccc}
\toprule
Method & 5.0k & 10.0k & Mean & SD \\
\midrule
TVDF & \textbf{0.742} & \underline{0.749} & \textbf{0.719} & 0.051 \\
TVMed & \underline{0.732} & 0.742 & 0.709 & 0.056 \\
TVDF$_{\text{VAR}}$ & 0.726 & \textbf{0.752} & 0.712 & 0.050 \\
TVCoverage & 0.729 & 0.748 & \underline{0.717} & 0.053 \\
\bottomrule
\end{tabular}
}
  \end{subtable}}
\end{table*}

We evaluate the impact of incorporating in-domain (ID) and/or out-of-domain (OOD) label information into BEACON’s sample selection process. The goal of this experiment is to determine whether relaxing the unsupervised assumption of TVDF leads to improved distribution alignment and downstream performance. Specifically, the TVDF (ID) variant uses label information from the in-domain data when constructing a domain-specific training set using TVDF for alignment. Correspondingly, TVDF (OOD) uses only out-of-domain label information, while TVDF (ID + OOD) incorporates both label sources to guide sampling. In each case, positive and negative samples are treated as separate distributions, and alignment is performed independently for each class. For each label-aware variant, we enforce a balanced (50/50) split between positive and negative samples, and compare performance against the standard unsupervised BEACON method.

Results in Table~\ref{tab:label_aware_macro} show that the unsupervised TVDF model slightly outperforms the label-aware variants overall. Specifically, the base TVDF model achieves the highest mean macro F1 score across budgets (0.716) and performs best at most individual budget levels (e.g., 5k and 10k). The TVDF (OOD) variant performs second best, achieving a mean macro F1 score of 0.700. A similar, though less pronounced, trend is observed for weighted F1 scores (Table~\ref{tab:label_aware_weighted}). The unsupervised TVDF model again achieves the highest average performance, with a weighted mean F1 score of 0.719. This is closely followed by the TVDF (OOD) and TVDF (ID) variants, with weighted mean F1 scores of 0.715 and 0.714, respectively. The smaller gains observed in the weighted setting suggest that label-aware sampling may negatively impact smaller or underrepresented domains by partitioning already limited data into even smaller class-specific distributions. In contrast, larger domains appear less sensitive to the addition of label information.

To further investigate these domain-level differences, we performed an additional offline experiment using a greedy selection strategy. In this setting, each domain selects whether to use the label information it has access to during sampling. For example, the TVDF (ID + OOD) variant compares the validation performance of all label configurations (no labels, ID labels, OOD labels, and both) and selects the best-performing option for a given domain. These results are included in our repository~\footnote{\url{https://github.com/nbpulsone/BEACON/blob/dist-alignment/greedy_la_results.pdf}} and show that label information can indeed improve upon the standard unsupervised model when selectively applied. Overall, these findings suggest that label information can provide valuable guidance for the EMAD sample selection process, though its effectiveness varies across domains.

\subsection{Domain Representations} \label{sub:representations}

We compare TVDF against variants using alternative domain representations. This experiment tests whether richer distribution representations improve alignment quality beyond centroid-based summaries. The alternative domain representations we test include medoids (TVMed), centroid+variance (TVDF$_{\text{VAR}}$), and embedding space coverage (TVCoverage). All methods follow a similar TVDF-based selection principle but differ in how distributions are represented, as discussed in Section~\ref{sec:distribution}.

Table~\ref{tab:representations_macro} shows that centroid-based TVDF performs best in the macro setting. Across all budget settings, TVDF achieves the highest average F1 score (0.716), with TVCoverage achieving a similar level of performance (0.711). While TVDF consistently performs best overall, performance at specific budget levels varies across methods. At smaller budgets, TVCoverage performs second best--for example, at the 5k budget it achieves an F1 score of 0.721, closely following TVDF (0.739). At larger budgets, TVMed performs second best, achieving an F1 score of 0.742 compared to TVDF’s 0.759.

The corresponding weighted results in Table~\ref{tab:representations_weighted} exhibit a similar trend. At lower budget settings and on average, TVDF achieves the highest performance. For example, at the 5k budget, TVDF attains an F1 score of 0.742, outperforming TVMed (0.732). At larger budgets (e.g., 10k), differences among the models are smaller. In this setting, TVDF$_{\text{VAR}}$ performs best with an F1 score of 0.752, slightly outperforming TVDF (0.749). On average, however, TVDF remains the top-performing method, with a weighted F1 score of 0.719.

These results suggest that simple centroid representations are often sufficient to capture the most useful structure of the embedding distributions. More complex representations may introduce additional noise 
without improving alignment, and therefore do not translate to better downstream performance.

\subsection{Domain-Agnostic Experiments} \label{sub:domain_agnostic}

We evaluate distribution-aware selection in a domain-agnostic setting, with the goal of isolating the effect of distribution alignment independent of domain structure. To accomplish this, we evaluate methods for downsampling the training data to 70\% of its original size. Experiments are conducted on a variety of datasets, including WDC Products, Amazon-Google (AG), Beers, and DBLP-ACM~\cite{mudgal2018deep, peeters2023wdc}. Results are depicted in Table~\ref{tab:domain_agnostic}.

\begin{table}[h]
    \centering
    \caption{Domain-Agnostic Results (F1 Score)}
    \resizebox{.75\linewidth}{!}{\begin{tabular}{lccccc}
\toprule
Method & WDC & AG & Beers & DBLP-ACM & Avg. \\
\midrule
Base (100\%) & \textbf{0.773} & \underline{0.697} & \textbf{0.519} & \textbf{0.970} & \textbf{0.740} \\
Random ($\rightarrow$70\%) & 0.750 & 0.712 & 0.431 & 0.953 & 0.712 \\
NC ($\rightarrow$70\%) & 0.428 & 0.185 & 0.379 & 0.304 & 0.324 \\
NN ($\rightarrow$70\%) & 0.741 & 0.628 & 0.381 & \underline{0.965} & 0.679 \\
TVDF ($\rightarrow$70\%) & \underline{0.762} & \textbf{0.727} & \underline{0.489} & \underline{0.965} & \underline{0.736} \\
\bottomrule
\end{tabular}}
    \label{tab:domain_agnostic}
\end{table}

Across most datasets, the model trained on the full dataset (BASE) achieves the best performance, suggesting that removing training data often harms generalization. The exception is Amazon-Google, where TVDF outperforms BASE (0.727 vs. 0.697 F1).

Among the downsampling methods, TVDF  outperforms other distribution-aware approaches. In contrast, Nearest to Centroid (NC) performs worst because it disproportionately retains negative (non-matching) samples while discarding many positive samples. Since negative samples dominate EM datasets and tend to occupy dense central regions of the embedding space, centroid-based selection favors these "typical" negatives, yielding a training set biased toward the negative class and substantially lower F1 scores.

Nearest Neighbor (NN) performs better than NC but does not outperform random sampling, suggesting that proximity-based selection alone can still introduce bias in the sampled distribution.

In contrast, TVDF accounts for how each candidate sample \emph{changes} alignment with the proxy distribution. This global perspective avoids over-concentrating samples and preserves broader coverage of the data distribution. As a result, TVDF achieves the strongest performance among downsampling methods, though it generally does not surpass training on the full dataset. These findings suggest that while removing training samples often reduces performance, distribution-aware selection can substantially mitigate this loss. More importantly, they highlight a key distinction between the standard downsampling setting and BEACON’s EMAD framework: BEACON uses distribution alignment to incorporate informative \emph{additional} out-of-domain samples under a fixed budget, whereas the domain-agnostic setting considered here is restricted to selecting a subset of the original training data. Nevertheless, when downsampling is necessary, distribution alignment remains an effective unsupervised strategy for reducing training data or annotation requirements while minimizing performance degradation.


\section{Conclusion}~\label{sec:conclusion}
In this paper we examined how label availability, domain representation, and domain-agnostic settings impact the performance of distribution alignment techniques in the BEACON pipeline. We found that incorporating label information--particularly for in-domain data--can help TVDF perform EM more effectively for certain domains. We also observed that using centroids for domain representation, despite their simplicity, is the most effective approach in practice. Finally, we showed that the distribution-alignment techniques used by TVDF can serve as an effective downsampling procedure, providing a strong alternative to random downsampling. 

Our study has limitations that motivate future work. We evaluate our methods using a single representative PLM (RoBERTa), and additional experiments with alternative PLMs could help assess the robustness of our findings across different embedding spaces. Further, while the label-aware and domain representation experiments are conducted on WDC, evaluating these methods on additional EM benchmarks may provide further insight into their generalizability, particularly beyond e-commerce applications. Overall, we hope that BEACON serves as a foundation for future research on distribution-aware learning in low-resource EM settings.

\begin{acks}
This material is based upon work supported by the National Science Foundation under Grant NRT-HDR-2021871. 
This work was supported in part by National Science Foundation (NSF) under award numbers NRT-HDR-2021871 and IIS-2348121 and by the United States-Israel Binational Science Foundation (BSF). Any opinions, findings, and conclusions or recommendations expressed in this material
are those of the author(s) and do not necessarily reflect the views of the National Science Foundation
\end{acks}

\bibliographystyle{ACM-Reference-Format}
\bibliography{sample}

\clearpage
\appendix

\begin{table*}[t]
    \centering
    \small
    \caption{\crvs{Toy EM dataset used to illustrate distribution-aware selection strategies.}}
    \label{tab:toy_em}
    \begin{tabular}{cllp{4.2cm}c}
\toprule
ID & Domain & Source A & Source B & Label \\
\midrule
$e_1$ & Computers &
Dell XPS 13 Laptop (16GB) &
16GB Dell XPS 13 Notebook&
Match \\

$e_2$ & Computers &
HP Pavilion 15 Laptop (8GB, i5) &
Lenovo IdeaPad 3 Laptop (8GB, i5) &
Non-match \\

$e_3$ & Computers &
Macbook Air 15 inch (M5) &
Macbook Pro 15 inch (M5) &
Non-match \\

$e_4$ & Electronics &
Sony WH-1000XM5 Wireless Headphones &
Sony WH1000XM5/Noise Cancelling  &
Match \\

$e_5$ & Electronics &
Apple AirPods Pro (2nd Gen) &
Samsung Galaxy Buds2 Pro &
Non-match \\

$e_6$ & Clothing &
Nike Men's Dri-FIT Shirt, Blue, Medium &
Nike Dri-FIT Tee for Men, Blue (M) &
Match \\

$e_7$ & Clothing &
Levi's 501 Original Jeans, Size 32 &
Wrangler Regular Fit Jeans, Size 32 &
Non-match \\
\bottomrule
\end{tabular}
\end{table*}

\crvs{\section{Illustrative Examples}\label{sec:appendix}}

In this section, we consider a toy example that illustrates how the concepts introduced in this paper are applied in practice. Table~\ref{tab:toy_em} consists of 7 candidate pairs from 3 domains, and will be used to demonstrate each of the concepts below.

\textbf{Label-Aware Sampling.} Consider training a model for the \emph{Computers} domain using the candidate pairs shown in Table~\ref{tab:toy_em}. Following Section~\ref{sec:distribution}, assume that label information is available for both the in-domain (ID) and out-of-domain (OOD) samples at selection time, and that we employ the label-aware variant of BEACON to determine which OOD samples to include in the training set. For this example, let the annotation budget be \textbf{5}. We first include all three in-domain samples in the training set, leaving a budget of \textbf{2} additional OOD samples.

Following the experimental setup described in Section~\ref{sec:experiments}, we partition the data into positive (matching) and negative (non-matching) distributions. As training begins, each candidate pair is converted into an embedding representation. We then compute the centroids of the in-domain positive and negative samples, which in this example correspond to the sets $\{e_1\}$ and $\{e_2,e_3\}$, respectively. Next, we evaluate each OOD candidate according to how much its inclusion improves the alignment between the corresponding training centroid and the validation centroid. In this toy example, adding $e_4$ produces the greatest improvement for the positive distribution, while adding $e_7$ produces the greatest improvement for the negative distribution. Consequently, $e_4$ and $e_7$ are selected and added to the training set, completing the sampling procedure.

\begin{figure}
    \centering
    \includegraphics[width=0.8\linewidth]{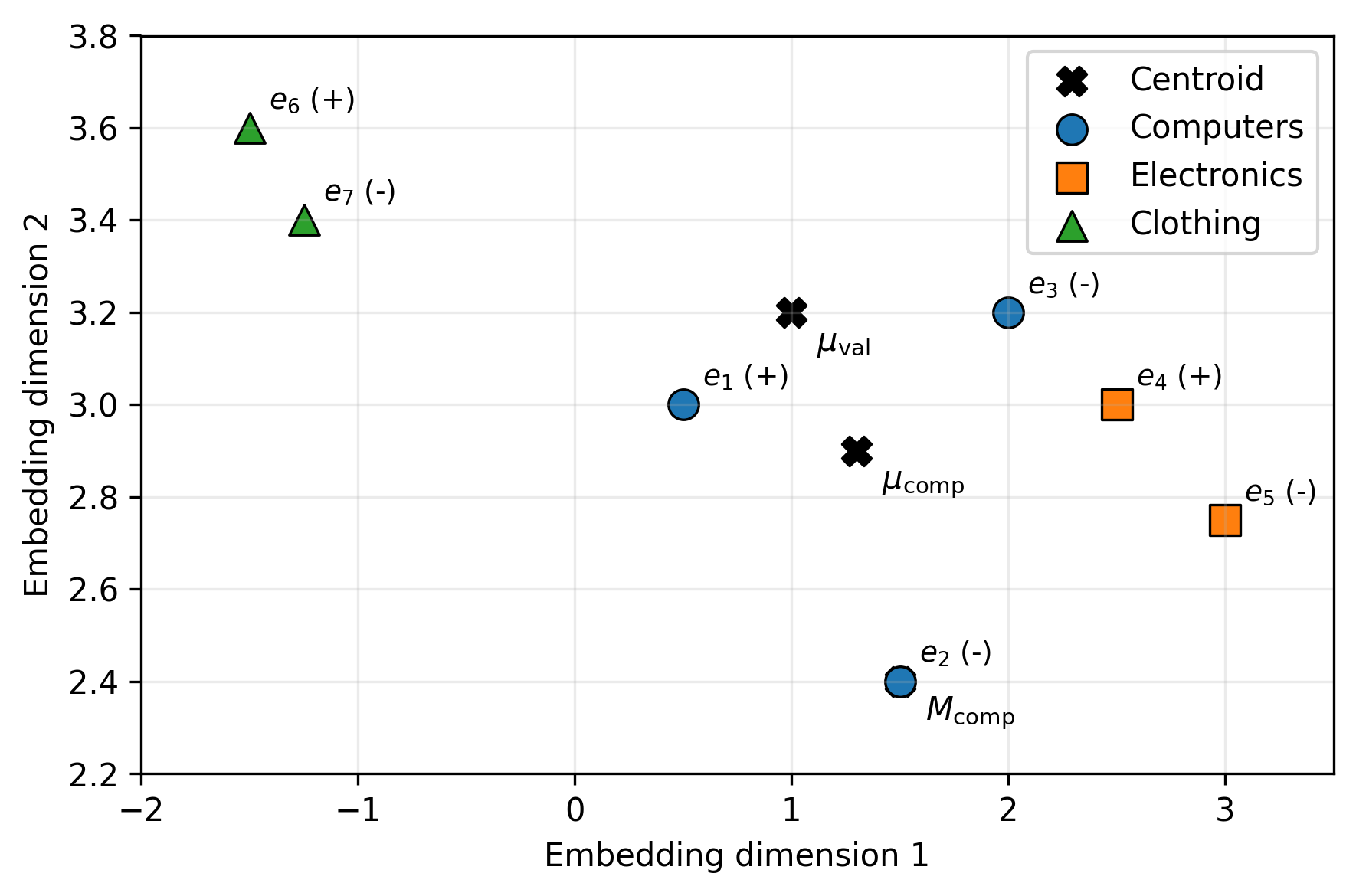}
    \caption{\crvs{The 2D embeddings for samples in Table~\ref{tab:toy_em}. Color and marker both denote product domain; +/- denotes label.}}
    \label{fig:toy_em}
\end{figure}

\textbf{Domain Representations.} Continuing with the goal of training an EM model for the Computers domain, consider the embedding representations obtained by converting each candidate pair in Table~\ref{tab:toy_em} into a high-dimensional embedding space using a PLM. Figure~\ref{fig:toy_em} depicts an illustrative two-dimensional projection of these embeddings. The figure also illustrates several representations of the Computers training distribution that can be used to guide sample selection, along with the proxy (validation) centroid, $\mu_{\text{val}}$, to which we wish to align the training distribution. For example, the centroid representation is given by $\mu_{\text{comp}}$, while the medoid representation is given by $M_{\text{comp}}$. We omit visualizations of the variance-based and coverage-based representations for simplicity.

BEACON uses the chosen representation to guide the sampling procedure. Under the centroid-based TVDF objective, candidate sample $e_7$ would be selected from the OOD pool because its inclusion shifts the training centroid $\mu_{\text{comp}}$ closer to the proxy centroid $\mu_{\text{val}}$. In contrast, under the medoid-based TVMed objective, sample $e_6$ could be selected because its inclusion changes the training medoid to a representation that is better aligned with the proxy distribution. This illustrates how different distribution representations can induce different notions of similarity and consequently lead to different sampling decisions.

\textbf{Domain-Agnostic Downsampling.} Finally, consider the case where we are no longer interested in training a model specifically for the Computers domain. Instead, our goal is to train a general EM model using approximately 70\% of the available training data. Furthermore, assume that label information is unavailable. For this example, we therefore seek to remove two samples from the dataset shown in Table~\ref{tab:toy_em}.

Consider the downsampling methods introduced in Section~\ref{sec:distribution}. The \textbf{BASE} method simply trains on all seven samples without performing any downsampling. One alternative is to \textbf{randomly} remove two samples from the dataset. For example, this procedure might remove $e_1$ and $e_4$, leaving only a single matching sample. More principled approaches use the embedding space to guide downsampling. The \textbf{NC} method retains the samples closest to the proxy centroid and therefore removes $e_6$ and $e_7$ in this example. Similarly, the \textbf{NN} method retains the samples nearest to any point in the proxy distribution, resulting in the removal of $e_1$ and $e_2$. Finally, the \textbf{TVDF} method removes the samples whose exclusion most improves alignment between the remaining training centroid and the proxy centroid. In this example, TVDF removes samples $e_2$ and $e_5$. This example illustrates that different downsampling strategies encode different notions of representativeness and can therefore produce substantially different training subsets.

\end{document}